\documentclass[aps,prl,twocolumn,superscriptaddress,showpacs]{revtex4}

\usepackage{graphicx}

\begin{document}


\title{Evidence for a change in the nuclear mass surface with the discovery of the most 
neutron-rich nuclei with $17\le Z\le 25$}


\author{O.B.~Tarasov}
\affiliation{National Superconducting
Cyclotron Laboratory, Michigan State University, East Lansing, MI,
USA 48824} \affiliation{Flerov Laboratory of Nuclear Reactions,
JINR, 141980 Dubna, Moscow Region, Russian Federation}
\author{D.J.~Morrissey}
\affiliation{National Superconducting Cyclotron Laboratory, Michigan
State University, East Lansing, MI, USA 48824} \affiliation{Dept.\
of Chemistry, Michigan State University, East Lansing, MI, USA
48824}
\author{A.M.~Amthor}
\affiliation{National Superconducting Cyclotron Laboratory, Michigan
State University, East Lansing, MI, USA 48824} \affiliation{Dept.\
of Physics and Astronomy, Michigan State University, East Lansing,
MI, USA 48824}
\author{T.~Baumann}
\affiliation{National Superconducting Cyclotron Laboratory, Michigan
State University, East Lansing, MI, USA 48824}
\author{D.~Bazin}
\affiliation{National Superconducting Cyclotron Laboratory, Michigan
State University, East Lansing, MI, USA 48824}
\author{A.~Gade}
\affiliation{National Superconducting Cyclotron Laboratory, Michigan
State University, East Lansing, MI, USA 48824} \affiliation{Dept.\
of Physics and Astronomy, Michigan State University, East Lansing,
MI, USA 48824}
\author{T.N.~Ginter}
\affiliation{National Superconducting Cyclotron Laboratory, Michigan
State University, East Lansing, MI, USA 48824}
\author{M.~Hausmann}
\affiliation{National Superconducting Cyclotron Laboratory, Michigan
State University, East Lansing, MI, USA 48824}
\author{N.~Inabe}
\affiliation{RIKEN Nishina Center, RIKEN, Wako-shi, Saitama 351-0198, Japan}
\author{T.~Kubo}
\affiliation{RIKEN Nishina Center, RIKEN, Wako-shi, Saitama 351-0198, Japan}
\author{A.~Nettleton}
\affiliation{National Superconducting Cyclotron Laboratory, Michigan
State University, East Lansing, MI, USA 48824} \affiliation{Dept.\
of Physics and Astronomy, Michigan State University, East Lansing,
MI, USA 48824}
\author{J.~Pereira} \affiliation{National Superconducting
Cyclotron Laboratory, Michigan State University, East Lansing, MI,
USA 48824}
\author{M.~Portillo}
\affiliation{National Superconducting Cyclotron Laboratory, Michigan
State University, East Lansing, MI, USA 48824}
\author{B.M.~Sherrill}
\affiliation{National Superconducting Cyclotron Laboratory, Michigan
State University, East Lansing, MI, USA 48824} \affiliation{Dept.\
of Physics and Astronomy, Michigan State University, East Lansing,
MI, USA 48824}
\author{A.~Stolz} \affiliation{National Superconducting
Cyclotron Laboratory, Michigan State University, East Lansing, MI,
USA 48824}
\author{M.~Thoennessen}
\affiliation{National Superconducting Cyclotron Laboratory, Michigan
State University, East Lansing, MI, USA 48824} \affiliation{Dept.\
of Physics and Astronomy, Michigan State University, East Lansing,
MI, USA 48824}

\date{\today}

\begin{abstract}
The results of measurements of the production of neutron-rich nuclei
by the fragmentation of a $^{76}$Ge beam are presented. The cross sections 
were measured for a large range of nuclei including fifteen new isotopes that 
are the most neutron-rich nuclides of the elements chlorine to manganese ($^{50}$Cl,
$^{53}$Ar, $^{55,56}$K, $^{57,58}$Ca, $^{59,60,61}$Sc, $^{62,63}$Ti, $^{65,66}$V,
$^{68}$Cr, $^{70}$Mn). The enhanced cross sections of several new nuclei relative 
to a simple thermal evaporation framework, previously shown to describe 
similar production cross sections, indicates that nuclei in the region around $^{62}$Ti 
might be more stable than predicted by current mass models and could be an 
indication of  
a new island of inversion similar to that centered on $^{31}$Na.
\end{abstract}

\pacs{27.50.+e,25.70.Mn}

\maketitle


Exploration of new isotopes with excess neutrons has yielded many surprises 
in nuclear physics. Weak binding of valence neutrons was found to result in 
an extended matter distribution or halo \cite{IT-JPG96}. The textbook shell 
structure of nuclei was found to evolve, primarily due to the importance of a 
tensor force in nuclei, unlike atomic shell structure \cite{TO-PRL07}. One 
of the earliest indications of significant changes in the structure of nuclei 
with excess neutrons was the discovery of enhanced nuclear binding of heavy 
sodium isotopes \cite{CT-PRC75}. This is now understood as a result of the 
significant contributions of intruder orbitals to the ground-state configuration 
of these isotopes \cite{XC-NPA75, EKW-PRC90}. The region of nuclei near 
$^{31}$Na where the neutron fp-shell contributes significantly to the 
ground-state structure is now known as the ``island of inversion.''

One of the challenges of nuclear physics is to push the study of neutron-rich 
isotopes to higher atomic number and an important benchmark in this work is 
to find the maximum number of neutrons that can be bound for each atomic number. 
Often the delineation of the heavy limit of stability, called the 
neutron drip-line,  itself has yielded surprises. 
Recently it was found that heavy isotopes of aluminum are likely more 
bound than predicted \cite{TB-N07}. Continuing this work, we report here 
the next step towards the fundamental goal of defining the absolute mass 
limit for chemical elements in the region of calcium.

The neutron drip line is only confirmed at present up to $Z=8$
($^{24}$O$_{16}$) through years of work at projectile fragmentation facilities
in France \cite{DGM-PRC90,OT-PL97}, the US \cite{MF-PRC96} and Japan \cite{HS-PLB99}.
The neutron drip line has been found to
rapidly shift to higher neutron numbers at $Z=9$, i.e., $^{31}$F$_{22}$ has
been observed several times \cite{MN-PLB02, AML-JPG02, EK-AIP07}.  
The nuclide $^{30}$F$_{21}$ has been shown not to exist and $^{32}$F$_{23}$ 
is thought to be unbound based on systematics while the particle stability 
of $^{33}$F$_{24}$ is an open question. The shift is predicted to continue at higher 
masses \cite{PH-N07, MS-PRC04, KTUY-PTP05} and
makes the search for the neutron drip line in this region even more challenging.
The fragmentation of $^{48}$Ca$_{28}$ projectiles at RIKEN in Japan \cite{MN-PLB02},
at GANIL in France \cite{AML-JPG02}, and at the NSCL in the US \cite{EK-AIP07, TB-N07, OT-PRC07}
has produced a number of heavier nuclei in this region including $^{40}$Mg$_{28}$
and $^{42}$Al$_{29}$, but no clear limit has been established yet.  On the other hand, all
nuclei up to $Z=12$ with $A=3Z+3$ have been shown to be unbound.  The fragmentation
of heavier stable beams such as $^{76}$Ge in which $^{52}$Ar 
was observed \cite{PFM-BAPS08} is necessary in order to go beyond the reported work. 

In the present work a primary beam of $^{76}$Ge was fragmented and a search
for new neutron-rich isotopes above $^{40}$Mg was carried out using the
recently developed tandem fragment separator technique \cite{TB-N07}.  It was
previously shown by Tarasov et al. \cite{OT-PRC07} that the cross sections for
projectile fragments in this region have an exponential dependance on $Q_g$ 
(the difference in mass-excess of the beam particle and the observed fragment that is  
independent of the target in contrast to the older $Q_{gg}$ four-body analysis 
applied to low energy reactions) and 
deviations from the predicted yield may be used to identify anomalies in the 
mass surface such as the new island of inversion near $^{62}$Ti predicted by 
Brown \cite{BAB-PPNP01} similar to the original island of inversion near $^{31}$Na.

A 132 MeV/u $^{76}$Ge beam from the Coupled Cyclotron
Facility at the National Superconducting Cyclotron Laboratory was
used to irradiate a series of $^{9}$Be targets (see below) and finally a tungsten target located
at the target position of the A1900 fragment separator \cite{DJM-NIMA03}.  The primary beam
current was monitored continuously and normalized to
Faraday cup readings during the course of the experiment.
The average beam intensity for the measurements of the most exotic
fragments was 32~pnA. The A1900 fragment separator was combined with the
S800 analysis beam line to form a two-stage separator system as described in 
Ref.~\cite{TB-N07}.  A two-stage separator provides
a high degree of rejection of unwanted reaction products and allows the 
identification of each fragment of interest. During the search for the most
exotic fragments, a Kapton wedge (20.2 mg/cm$^2$) was used at the center
of the A1900 to reject less exotic fragments at the A1900 focal plane by an
8~mm aperture.  The transmitted fragments
passed on to the S800 beam line for event-by-event momentum analysis and particle
identification. The momentum acceptance of the A1900 was set to $\Delta p/p = \pm$0.05\%,
$\pm$0.5\%, $\pm$1\% and $\pm$2.5\% as the production rate of the increasingly exotic
nuclei decreased, always with an angular acceptance of 8.2~msr. 

\begin{figure}[b]
\includegraphics[width=0.75\textwidth]{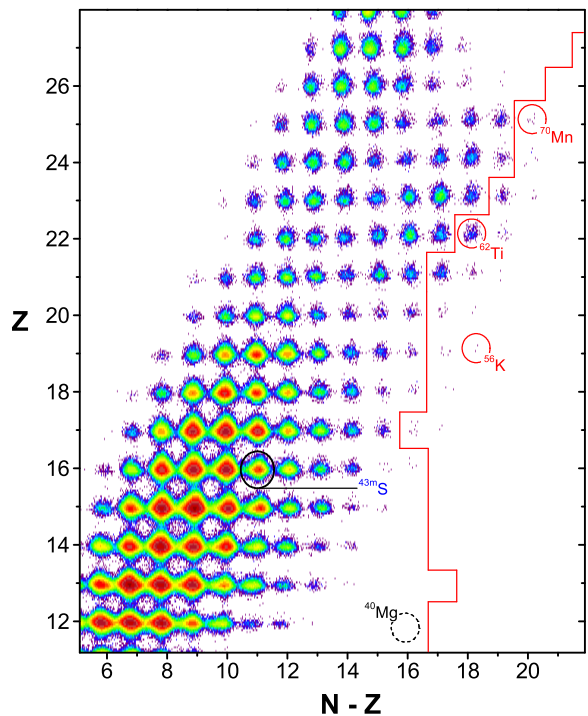}%
\caption{(Color online) Particle identification plot of the measured 
atomic number, $Z$, versus the calculated function $N-Z$ for the nuclei 
observed in this work, see the text for details. The limit of previously 
observed nuclei is shown by the solid line as well as the positions of several 
reference nuclei.\label{pid}}
\end{figure}

The particles of interest 
were stopped in a stack of eight silicon PIN diodes (50$\times$50~mm$^2$)
with a total thickness of 8.0~mm that provided multiple measurements of the
energy-loss and thus made redundant determinations of the nuclear charge
of each fragment along with the total kinetic energy.  The time of flight (TOF) of
each particle that reached the detector stack was measured in four ways:
(1) over the 46.04~m flight path between a thin plastic scintillator 
located at A1900 focal plane and the second PIN detector, (2) over the 20.97~m flight
path between another thin plastic scintillator at the object
point of the S800 analysis beam line and the third PIN detector, (3) over
the entire 81.51~m flight path by measuring the arrival time relative to the
phase of the cyclotron rf-signal and the third PIN detector, and (4) over the 25.07~m 
path between the scintillators at the object point and the A1900 focal plane. The 
magnetic rigidity of 
each particle ($B\rho$), which is proportional to the momentum/charge, was
obtained by combining a position measurement in two parallel-plate
avalanche counters (PPAC's) at the dispersive plane of the S800 analysis line
with NMR measurements of the dipole fields. The simultaneous measurement
of multiple $\Delta E$ signals, the magnetic rigidity, the total
energy, and the TOF's of each particle provided an
unambiguous identification of the atomic number, charge state and
mass number of each isotope.  The position in a PPAC in front of the silicon
detectors allowed additional discrimination against various scattered particles. 
The detection system and particle identification was calibrated with the 
primary beam and by the locations of gaps corresponding to unbound nuclei 
in the particle identification spectrum.


\begin{figure*}  
\includegraphics[width=0.85\textwidth]{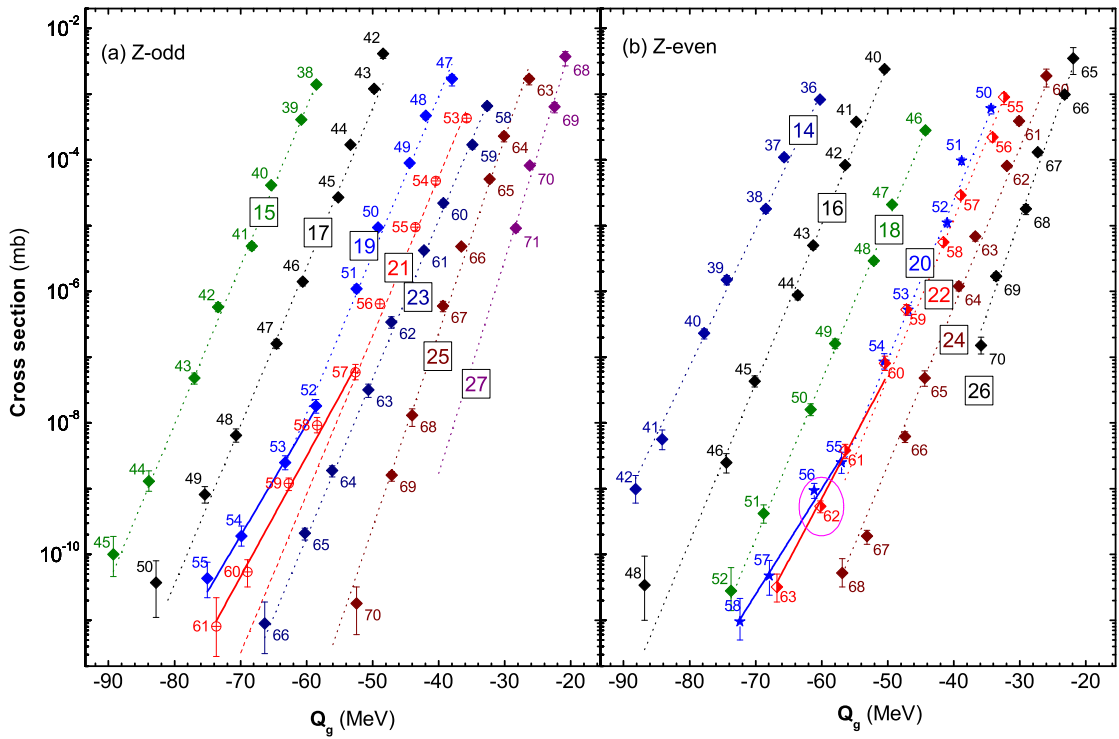}%
\caption{(Color online) The cross sections for production of neutron-rich nuclei with
(a) odd atomic numbers and (b) even atomic numbers with a Be target in the 
present work. See text for explanation of the lines. The cross section for $^{62}$Ti at 
the center of the proposed new island of inversion~\protect{\cite{BAB-PPNP01}} 
is circled.\label{Fig_cs-Be}}
\end{figure*}

The approach of changing the target thickness while maintaining the magnetic 
system at essentially constant magnetic rigidities was used in order to simplify
the particle identification during the run while mapping the (parallel) momentum
distributions of the less exotic fragmentation products.  The details will be described in 
a subsequent report \cite{OT-prep09}. Beryllium targets of thicknesses 9.8, 97.5, 145, 191, 
288, 404, and finally 629 mg/cm$^2$ were inserted sequentially over
the course of the experiment.  The particle identification depended critically on the magnetic 
rigidity of the S800 analysis line and thus was constant during the experiment whereas the 
average energy (and rigidity) of a given reaction product decreased dramatically with 
increasing target thickness.  Thus, the isotopic selection of this system moved towards 
heavier fragments with increasing target thickness.
The magnetic rigidities were $B\rho =4.355$ Tm (after the target), 4.334~Tm (after the wedge), 
4.320~Tm (after the first TOF scintillator) and 4.305~Tm (after the second TOF scintillator) for the
629~mg/cm$^2$ target when the most exotic nuclei were observed. The power of
this approach can be seen in Fig.~\ref{pid}\ that shows the distribution of fully-stripped 
reaction products observed during this work. The range of fragments is shown as the 
measured atomic number, $Z$, versus 
the function $N-Z$ where $N$ is the calculated neutron number. The absolute identification 
of the individual isotopes in Fig.~\ref{pid} was confirmed 
by the identification of gamma rays from isomeric states in  $^{32}$Al and $^{43}$S with 
a high-purity germanium detector placed near the silicon detector stack. 

A search for the most exotic nuclei was carried out by performing several runs
with the 404 and 629 mg/cm$^2$ Be targets for a total of 50.83 hours and additional
runs with a 567 mg/cm$^2$ $^{nat}$W target with various stripper foils
for a total of 77.66 hours. In contrast to previous studies, i.e.~\cite{MF-PRC96, OT-PRC07},
the present efficiency was dominated by the acceptance
of the tandem separator and not by the data acquisition system due to the high selectivity 
of the separator system.  The angular and total 
transmissions were calculated for each isotope for each run using a model of the momentum 
distribution with parameters obtained from the measured parallel momentum distributions.  For 
example, the angular and total transmissions for $^{66}$V with the 404~mg/cm$^2$ target 
were 99.0$\strut^{+0.7}_{-1.6}$\% and 58$\pm 4$\%, respectively. The typical uncertainty 
in the total transmission for other cases was $\approx$5\%.

\begin{figure}
\includegraphics[width=0.35\textwidth]{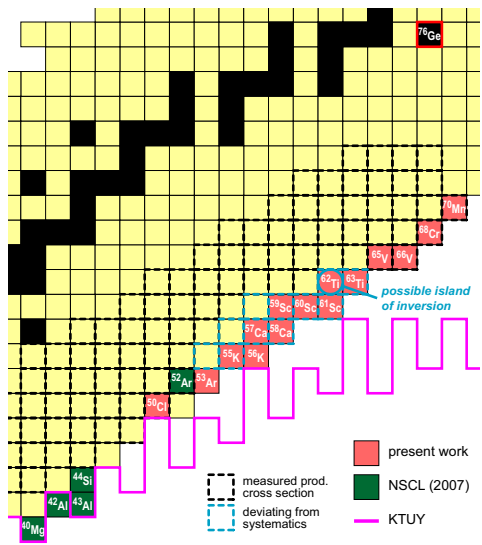}%
\caption{(Color online) The region of the chart of nuclides under investigation. The solid line is the 
limit of bound nuclei from the KTUY mass model \protect{\cite{KTUY-PTP05}}.  Nuclei
in the green squares were recently discovered \protect{\cite{OT-PRC07,TB-N07,PFM-BAPS08}}, 
those in red squares are new in this work, and the cross sections for those in dashed boxes 
are shown in Fig.~\protect{\ref{Fig_cs-Be}}. The center of the new proposed island of 
inversion at $^{62}$Ti \protect{\cite{BAB-PPNP01}} is highlighted.  \label{chart}}
\end{figure}

The observed fragments include fifteen new isotopes that are the most neutron-rich nuclides 
of the elements chlorine to manganese ($^{50}$Cl,
$^{53}$Ar, $^{55,56}$K, $^{57,58}$Ca, $^{59,60,61}$Sc, $^{62,63}$Ti, $^{65,66}$V,
$^{68}$Cr, $^{70}$Mn). The new neutron-rich nuclei observed in this work are those 
events to the right of the solid line in Fig.~\ref{pid} and are indicated by the red squares 
in Fig.~\ref{chart}. We should note that the observation of  $^{51}$Cl was reported 
some time ago among the products from the reaction of
$\rm ^{48}Ca + ^{64}Ni$ at the relatively low energy of 44 MeV/u \cite{ML-ZPA90}.  While
not conclusive, the previous identification of this isotope may have been masked by the presence
of the hydrogen-like ion $^{48}$Cl$^{16+}$ produced at the same time.


The cross sections for the production of all of the nuclei observed with the beryllium targets
 are shown in Figs.~\ref{Fig_cs-Be} (a) and (b) for product nuclei with
 odd and even atomic numbers, respectively. No nuclear reaction model can
 reproduce the very low yields of the exotic nuclei observed in this study and thus we
 will consider the general behavior of the production cross sections.  Projectile
 fragmentation processes are usually described as a sudden process that forms
 an excited prefragment followed by statistical decay.  Charity \cite{RJC-PRC98} has
 pointed out that the sequential evaporation of light particles from sufficiently excited
 nuclei follows a general pattern that leads to a somewhat uniform distribution of final
 products. Such uniform distributions underlay the EPAX systematics \cite{KS-PRC00}
 that is often used to predict the yields of fragmentation reactions when designing experiments.
We have recently shown \cite{OT-PRC07} that the yields of neutron-rich projectile fragments
generally show a smooth exponential decline with the function:
$$ Q_g = ME(Z_p,A_p)-ME(Z,A)$$
where $ME(Z,A)$ is the mass excess of a product with atomic number, $Z$, and mass
number, $A$, $Z_p=32$ and $A_p=76$. The $Q_g$ function depends on the relative binding
energies of the projectile fragments without regard to the target nucleus and is a
plausible basis for comparison of products from a process that creates a set of
highly excited intermediate nuclei that then statistically populate the mass surface.

Figure \ref{Fig_cs-Be} shows that the $Q_g$ function using masses from from 
Ref.~\cite{KTUY-PTP05} (a representative calculation selected from among many others 
that give similar results)  provides an excellent systematization of the 
new data with the logarithm of the cross sections for each isotopic chain falling on an approximately 
straight line. The isotopic chains with $15\le Z \le 24$ can be fit with a single exponential slope 
of $1/1.8 MeV$, as shown by the dashed lines.  However, closer inspection shows that the 
heaviest members of the isotopic chains with $Z$=19, 20, 21 and 22 break away from the 
uniform fit and the heaviest four or five isotopes have a shallower slope or enhanced cross 
sections.  Recall that the masses of these most neutron-rich nuclei are not fit in the model but 
are extrapolated. This might indicate that these nuclei are
more bound (i.e., less negative $Q_g$) than current mass models predict. One reason for 
a stronger binding can be deformation. In a shell-model framework, the wave
functions of the ground and low-lying excited states of nuclei in the new 
island of inversion around $^{62}$Ti would be dominated by neutron
particle-hole intruder excitations across the $N=40$ sub-shell gap,
leading to deformation and shape coexistence.


In summary, fifteen new neutron-rich isotopes were observed by 
fragmentation of a $^{76}$Ge beam provided evidence for  The general decline of the cross sections for 
the production of all of the observed neutron-rich isotopes with increasing mass
number is consistent with the well-known EPAX parameterization
and with the recently established $Q_g$ systematics.  The new data show 
a smooth exponential dependence with the mass excess of the observed
fragment with a few exceptions.  The fact that the cross sections of the most
neutron-rich nuclei with $Z=19$ to 22 are enhanced relative to the lighter isotopes   
may indicate that these nuclei and their precursors are more bound than predicted 
and is reminiscent of the discovery of the island of inversion at $N=20$.

The authors would like to acknowledge the work of the operations
staff of the NSCL to develop the intense $^{76}$Ge beam necessary
for this study. This work was supported by the U.S.~National
Science Foundation under grant PHY-06-06007.



\end{document}